\documentclass[aps,prx,twocolumn,amsmath,amssymb,superscriptaddress,longbibliography]{revtex4-2}
\usepackage[breaklinks,colorlinks,bookmarks=false,citecolor=blue,linkcolor=red,urlcolor=blue]{hyperref}
\usepackage{blindtext,enumitem,graphicx,dcolumn,color,xcolor,amsmath,braket,bm,changes,chemformula,cleveref,times}

\graphicspath{{figures/}}
\usepackage{blindtext}
\usepackage{changes}

\usepackage{graphicx}
\usepackage{tikz}
\usepackage{pgfplots}
\usepgfplotslibrary{groupplots}
\usepackage{chemformula}

\definecolor{darkpastelgreen}{rgb}{0.01, 0.75, 0.24}

\newcommand*\chem[1]{\ensuremath{\mathrm{#1}}}
\newcommand{\mycomment}[1]{}

\crefname{equation}{Eq.}{Eqs.}
\Crefname{equation}{Eq.}{Eqs.}
\crefname{figure}{Fig.}{Figs.}
\Crefname{figure}{Fig.}{Figs.}
\crefname{section}{Sec.}{Secs.}
\Crefname{section}{Sec.}{Secs.}

\begin{document}
\title{Fragmented superconductivity in the Hubbard model as solitons in Ginzburg-Landau theory}

\author{Niccol\`o Baldelli}
\email{niccolo.baldelli@icfo.eu}
\affiliation{ICFO - Institut de Ci\`encies Fot\`oniques, The Barcelona Institute of Science and Technology, 08860 Castelldefels (Barcelona), Spain}
\affiliation{Center for Computational Quantum Physics, Flatiron Institute, 162 5th Avenue, New York, New York 10010, USA}

\author{Hannes Karlsson}
\affiliation{Max Planck Institute for the Physics of Complex Systems, N\"othnitzer Strasse 38, Dresden 01187, Germany}

\author{Benedikt Kloss}
\author{Matthew Fishman}
\affiliation{Center for Computational Quantum Physics, Flatiron Institute, 162 5th Avenue, New York, New York 10010, USA}

\author{Alexander Wietek}
\email{awietek@pks.mpg.de}
\affiliation{Max Planck Institute for the Physics of Complex Systems, N\"othnitzer Strasse 38, Dresden 01187, Germany}

\begin{abstract}
The phenomena of superconductivity and charge density waves are observed in close vicinity in many strongly correlated materials. Increasing evidence from experiments and numerical simulations suggests both phenomena can also occur in an intertwined manner, where the superconducting order parameter is coupled to the electronic density. Employing density matrix renormalization group simulations, we investigate the nature of such an intertwined state of matter stabilized in the phase diagram of the elementary $t$-$t^\prime$-$U$ Hubbard model in the strong coupling regime. Remarkably, the condensate of Cooper pairs is shown to be fragmented in the presence of a charge density wave where more than one pairing wave function is macroscopically occupied. Moreover, we provide conclusive evidence that the macroscopic wave functions of the superconducting fragments are well-described by soliton solutions of a Ginzburg-Landau equation in a periodic potential constituted by the charge density wave. In the presence of an orbital magnetic field, the order parameters are gauge invariant, and superconducting vortices are pinned between the stripes. This intertwined Ginzburg-Landau theory is proposed as an effective low-energy description of the stripe fragmented superconductor. 
\end{abstract}

\date{\today}
\maketitle

\section{Introduction}
Intertwined orders come in many guises in strongly correlated electron systems~\cite{Fradkin2015,Fernandez2019,Vojta2009}. Continuous and discrete symmetries are broken in a manner in which order parameters transform genuinely non-trivially in the entire symmetry group. A prominent example is constituted by stripe order, where both the translational and spin rotation symmetry are broken to a state which intertwines a charge density wave (CDW) and a spin density wave. Initially proposed in Hartree-Fock studies~\cite{Zaanen1989,Poilblanc1989,Machida1989,Kato1990}, modern numerical techniques have in the last years succeeded in firmly establishing it as the ground state in certain parameter regimes of the strongly correlated repulsive Hubbard model in two spatial dimensions~\cite{LeBlanc2015,Zheng2017,Huang2017,Huang2018,Qin2020}. As such, these orders are relevant to the physics of high-temperature superconductivity where electron-electron interactions in the form of the Hubbard model are understood to play a crucial and likely decisive role~\cite{Anderson1987,Zhang1988,Emery1988}. 

Historically, the competition between superconductivity and stripe order has been discussed thoroughly~\cite{Kivelson2003,Wen2019}. Both stripe order and superconductivity have by now been found to be realized in different parameter regimes of the paradigmatic Hubbard and $t$-$J$ models~\cite{LeBlanc2015,Zheng2017,Huang2017,Huang2018,Qin2020,Mai2022,Mai2023,Huang2023,Xu2022,Xiao2023,Simkovic2022,Wietek2022t,Wietek2021}, and have also been found when further neighbor hopping processes are included ~\cite{Gong2021,Jiang2021,HCJiang2021}. Moreover, numerical studies  in the last years have discovered that superconductivity can also be intertwined with CDW order~\cite{Gong2021,Jiang2021,HCJiang2021,Jiang2023,Jiang2023b,Xu2023,Lu2023,Wang2022,Ponsioen2023}. A particular form of intertwined CDW order and superconductivity is the so-called pair-density wave state~\cite{Agterberg2008,Dai2018,Agterberg2020,Huang2022,Jiang2023b,Setty2023,Wang2018}, for which recently evidence has been reported in several unconventional superconductors~\cite{Hamidan2016,Gu2023}. The coexistence of CDW order, which breaks translational symmetry, and superconductivity is sometimes also referred to as supersolid order. Previously, one of the authors reported that the CDW can ``fragment'' the superconducting condensate~\cite{Wietek2022}. Fragmentation refers to the phenomenon when fermion pairs (or bosons in Bose-Einstein condensates) condense into not just one, but multiple macroscopic wave functions. The ground state in a certain regime of the $t$-$t^\prime$-$J$ model exhibits exactly one fragment per charge stripe of the superconducting condensate, interpreted as an emergent array of coupled Josephson junctions.

In this manuscript, we strengthen our understanding of this intriguing form of order. Employing density matrix renormalization group (DMRG)~\cite{White1992,White1993,Schollwock2005,Schollwock2011} simulations, we show the fragmented superconducting stripe order is also realized in the paradigmatic Hubbard model on the square lattice  for systems up to width $W=6$, both with open and cylindrical boundary conditions. Additionally, we investigate the effect of an orbital magnetic field on the fragments of the Cooper condensate. Moreover, we propose a minimalistic macroscopic model to describe the superconducting condensates which is given by an intertwined Ginzburg-Landau theory~\cite{Ginzburg1950}. Importantly, the mass term is chosen to be position dependent and exactly proportional to the hole-density modulation of the CDW. A one-to-one comparison between the solution of the Ginzburg-Landau equation and the numerical data from DMRG is performed and yields detailed agreement. The fragmentation of the superconducting condensate is then understood as a quantum tunneling between several soliton solutions of the Ginzburg-Landau equation.

\begin{figure}
    \centering
    \includegraphics[width =\columnwidth]{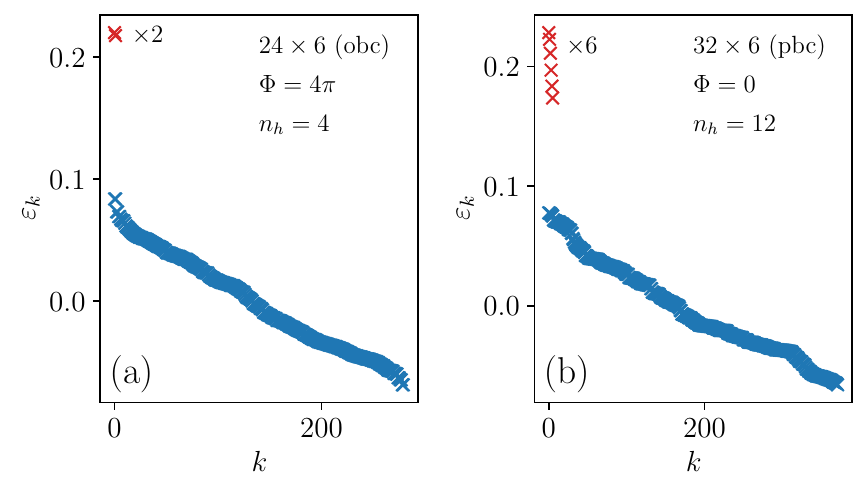}
    \vspace{-0.7cm}
    \caption{Fragmentation of the Cooper condensate apparent in the spectrum of the singlet two-particle density matrix of the ground state of the Hubbard model at small doping for $t^\prime/t=0.2$ and $U/t = 10$. (a) $24 \times 6$ square lattice (open boundary conditions) doped with $n_h=4$ holes away from half-filling and total magnetic flux $\Phi=4\pi$ through the full sample. Two dominant eigenvalues are observed. (b) $32 \times 6$ cylinder doped with $n_h=12$ holes without magnetic flux. We observe six dominant eigenvalues. In both cases, the number of dominant eigenvalues exactly matches the number of CDW maxima.}
    \label{fig:spectra}
\end{figure}
\begin{figure*}
\includegraphics[width=\textwidth]{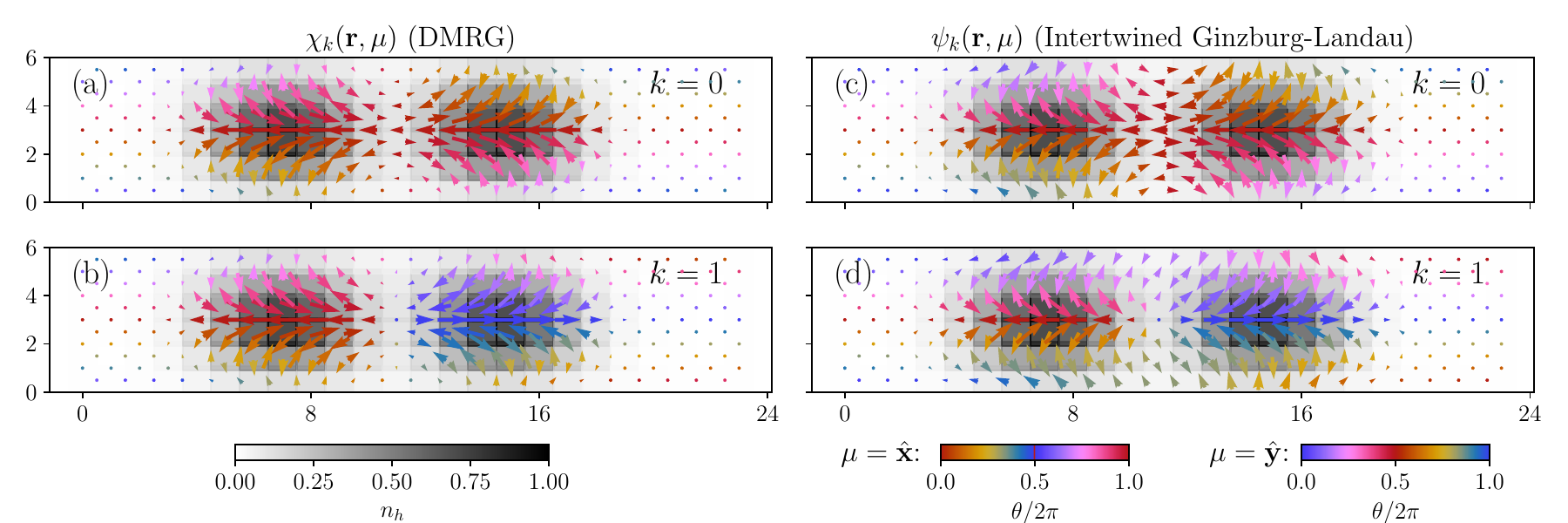}
\caption{Comparison between macroscopic wave functions $\chi_k(\bm{r}, \mu)$ from DMRG simulations and $\psi_k(\bm{r}, \mu)$ from intertwined Ginzburg-Landau theory.
(a,b) Macroscopic wave functions of ground state of the $t$-$t^\prime$-$U$ Hubbard model from DMRG with total magnetic flux $\Phi = 4\pi$ on a $24 \times 6$ lattice with open boundary conditions when doped with $n_h=4$ holes away from half-filling. Here we take $t^\prime / t = 0.2$, $U/t=10$. (a) (resp. (b)) shows the values of $\chi_k(\bm{r}, \mu)$ for the $k=0$ (resp. $k=1$) fragment of the condensate. The length of the arrows is proportional to $|\chi_k(\bm{r}, \mu)|$ and the color indicates the complex angle of $\chi_k(\bm{r}, \mu)$, where for $\mu=\hat{\bm{y}}$ we shift the phase by $\pi$, to reflect the $d$-wave nature of the order parameter. The grayscale of the background squares shows the value of the local hole density $n_h(\bm{r})$. (c,d) Both soliton solutions $\psi_k(\bm{r}, \mu)$ ($k=0,1$) of the intertwined Ginzburg-Landau theory for $\alpha(\bm{r}) = \alpha n_h(\bm{r})$, where $n_h(\bm{r})$ is taken from the values of the DMRG ground state and the free parameters $\alpha=-2.331$, $\beta=14.465$ are found by matching the DMRG data to Ginzburg-Landau theory. We observe close agreement between both $\chi_k(\bm{r}, \mu)$ and $\psi_k(\bm{r}, \mu)$.}
\label{fig:wfs_hfield}
\end{figure*}

\section{Model and observables}
We study the $t$-$t^\prime$-$U$ Hubbard model on a square lattice defined by the Hamiltonian
\begin{equation}
    H = -\sum_{ij,\sigma}\left(t_{ij}c^{\dagger}_{i\sigma} c_{j\sigma} + \textrm{h.c.}\right) 
    + U \sum_{i}n_{i\uparrow}n_{i\downarrow},
    \label{hubbard_ham}
\end{equation}
where $t_{ij} = t$ for nearest neighbor lattice sites $i$ and $j$ and $t_{ij} = t^\prime$ for next-nearest neighbor sites. Here, $c^{\dagger}_{i\sigma}, c_{i\sigma}$ denote the fermionic creation and annihilation operators of spin $\sigma=\uparrow, \downarrow$ and $n_{i\sigma} = c^{\dagger}_{i\sigma} c_{i\sigma}$ denotes the number operator.
We additionally study the effect of a uniform orbital magnetic field $B = \nabla \times \bm{A}$, orthogonal to the plane of the lattice. We employ the Landau gauge
\begin{equation}
    \bm{A}(\bm{r}) = (A_x, A_y) = (0, 2\pi\phi x),
\end{equation}
where $\bm{r}=(x,y)$ denotes a real space coordinate. The total magnetic flux through the simulation cluster is denoted by $\Phi = \int_S \textrm{d}\bm{S}\; B$. The Landau gauge is well suited for open and cylindrical boundary conditions, since in the latter case it remains translationally invariant in the $y$-direction. The coupling of the model to the background magnetic field is implemented via a Peierls substitution
\begin{equation}
    t_{ij} \rightarrow t_{ij}
    \exp\left[-i \frac{e}{\hbar}\int_{\bm{r}_i}^{\bm{r}_j} \textrm{d}\bm{r} \cdot \bm{A}(\bm{r})  \right],
\end{equation}
where $e$ denotes the elementary charge. In the following, we set $e=1$, $\hbar=1$ and focus on the model parameters $U/t = 10$ and $t'/t = 0.2$. For our geometry, the Peierls substitution reduces to $t\rightarrow te^{i2\pi\phi x}$ in the $y$-direction, and $t'\rightarrow t'e^{i2\pi\phi (x+1/2)}$ as explained in Ref.~\cite{Hatsugai1990}. 

In large orbital magnetic fields, this model has been studied in the context of cold-atoms experiments~\cite{Cocks2012,Stenzel2020,Palm_2023}. As worked out extensively in Ref.~\cite{Jiang2023} for $U/t=12$, on six-leg cylinders this set of parameters stabilizes a superconducting state coexisting with a CDW at small hole-doping. Analogously, the coexistence of CDW and superconducting order has been reported at larger hole-doping by a combination of auxiliary-field quantum Monte Carlo and DMRG~\cite{Xu2023}. Moreover, the phase diagram at small hole-doping has been found to be in close agreement with the $t$-$t^\prime$-$J$ model, where similarly a coexisting CDW and superconducting state have been reported~\cite{Jiang2020,Jiang2021,Gong2021,Wietek2022}. We would like to emphasize that our findings are in full agreement with these previous studies. 

Our aim is to elucidate the relation between charge and pairing degrees of freedom. We consider the hole density
\begin{equation}
n_h(\bm{r}) = 1 - n(\bm{r}),
\end{equation}
where $\bm{r}$ denotes the lattice position and $n(\bm{r})$ the electron density. 
As the pairing mechanism in the lightly-doped Hubbard model has been diagnosed in previous studies to be of singlet-pairing type, we consider
\begin{equation}
\label{eq:singletpairingmatrix}
\rho_S(\bm{r}_i, \bm{r}_j | \bm{r}_k ,\bm{r}_l) = 
\langle \Delta_{\bm{r}_i\bm{r}_j}^\dagger \Delta_{\bm{r}_k\bm{r}_l}\rangle,
\end{equation}
where the singlet-pairing operators $\Delta_{\bm{r}_i\bm{r}_j}$ are defined as
\begin{equation}
\label{eq:singletpairingop}
\Delta_{\bm{r}_i\bm{r}_j}^\dagger= \frac{1}{\sqrt{2}}
\left( 
c^\dagger_{i\uparrow} c^\dagger_{j\downarrow} - 
c^\dagger_{i\downarrow} c^\dagger_{j\uparrow}
\right).
\end{equation}
Moreover, as the pairing in this superconducting phase was established as local in real space~\cite{Jiang2023,Xu2023,Wietek2022,HCJiang2021,Jiang2021,Jiang2020,Gong2021}, we restrict ourselves to investigating the nearest-neighbor singlet-pairing density matrix
\begin{equation}
\label{eq:nnsingletpairingmatrix}
\rho_S(\bm{r}_i, \mu | \bm{r}_j, \nu) = \rho_S(\bm{r}_i ,(\bm{r}_i + \mu) | \bm{r}_j, (\bm{r}_j + \nu)),
\end{equation}
where $\mu= \hat{\bm{x}},  \hat{\bm{y}}$ (resp. $\nu$) denote the vectors connecting nearest-neighbors on the square
lattice. We exclude site-local contributions from density and spin correlations by setting $\rho_S(\bm{r}_i, \mu | \bm{r}_j, \nu) =0$ whenever the two links $(\bm{r}_i, \bm{r}_i +\mu)$ and $(\bm{r}_j, \bm{r}_j +\nu)$ are overlapping, see Eq.~(8) in Ref.~\cite{Wietek2022}. Because of this,$\rho_S(\bm{r}_i, \mu | \bm{r}_j, \nu)$ ceases to be positive definite and therefore can have negative eigenvalues. Off-diagonal long-range order occurs whenever 
\begin{equation}
    \rho_S(\bm{r}_i, \mu | \bm{r}_j, \nu) \rightarrow C \neq 0 \quad\textrm{ for } \quad|\bm{r}_i - \bm{r}_j| \rightarrow \infty.
\end{equation}
However, the information contained in $\rho_S(\bm{r}_i, \mu | \bm{r}_j, \nu)$ in \cref{eq:nnsingletpairingmatrix} is richer and allows for additional insights beyond diagnosing long-range order. As a Hermitian matrix, $\rho_S(\bm{r}_i, \mu | \bm{r}_j, \nu)$ has an eigendecomposition with real eigenvalues $\varepsilon_n$, 
\begin{equation}
\label{eq:nnsingletpairingmatrixdecomp}
\rho_S(\bm{r}_i, \mu | \bm{r}_j, \nu) = 
\sum_n \varepsilon_n \chi_n^*(\bm{r}_i, \mu) \chi_n(\bm{r}_j, \nu).
\end{equation}
The onset of Cooper pair condensation can then be inferred by the presence of either one or multiple dominant eigenvalues~\cite{Legett2006,Wietek2022,Penrose1956}. This criterion is also known as the Penrose-Onsager criterion \cite{Penrose1956} of superconductivity. In case a single dominant eigenvalue $\varepsilon_0$ is observed the condensate is called simple. Otherwise, if multiple $\varepsilon_k$ are dominant, the condensate is said to be \textit{fragmented}. The dominant eigenvalues $\varepsilon_k$ are referred to as the condensate fractions, while the corresponding eigenvectors $\chi_k(\bm{r}, \mu)$ are referred to as macroscopic wave functions. Notice, that the eigenvectors depend on the position of the lattice $\bm{r}$ as well as the orientation of the bond $\mu$ and are, hence, defined on the bonds of the lattice. 

%Under a local gauge transformation,
%\begin{equation}
%\label{eq:gaugetrafo}
%c^\dagger_{\bm{r}_i} \rightarrow  e^{i\phi(\bm{r}_i)} c^\dagger_{\bm{r}_i}, \quad A(\bm{r}) \rightarrow A(\bm{r}) - \nabla\phi(\bm{r}),
%\end{equation}
%the macroscopic wave functions transform as,
%\begin{equation}
%\chi_k(\bm{r}_i, \mu) \rightarrow e^{2i\phi(\bm{r}_i)}\chi_k(\bm{r}_i, \mu).
%\end{equation}

\section{Ground state properties from DMRG}

To investigate the onset of intertwined order in the Hubbard model we perform DMRG calculations to obtain the ground state properties of the system. In this scheme, the 2D system is mapped to a 1D chain suitable for DMRG, with the drawback of the $x$-direction hopping term in the Hamiltonian becoming long-range. This leads to an exponential increase in the computational cost when increasing the width $W$ of the system, limiting the simulations to narrow cylinders. For this reason, in this work, we choose $W = 6$. 

We obtain results for systems of length $L = 24 $ with open boundary conditions and $L = 32$ with periodic boundary conditions in the $y$-direction. On a cylindrical geometry the  Hamiltonian in Eq.~(\ref{hubbard_ham}) is not gauge invariant as a gauge transformation introduces a magnetic flux piercing the cylinder, corresponding to a ground state with a finite momentum $k_y$ along the $y$-direction. As this is unphysical when looking for ground state properties we fix the gauge in order to stay in a sector with zero net $k_y$ momentum. Analogously to previous works \cite{Gong2021,Xu2023} we restrict to the $S^z_{tot} \equiv \langle\sum_i  n_{i\uparrow} - n_{i\downarrow}\rangle = 0$ sector of total spin projection in the $z$ direction. Exploiting this symmetry, together with the conservation of the total charge $N \equiv \langle\sum_i  n_{i\uparrow} + n_{i\downarrow}\rangle$ drastically reduces the computational load, allowing us to access larger bond dimensions. For the initial state of DMRG, we choose a physically motivated Néel product state with the holes evenly distributed in pairs along the $x$ direction. Special care had to be taken to make sure that this choice of the initial state does not lead to local minima of the DMRG optimization. As the convergence properties are highly dependent on the initial ansatz, we also ran the calculations with different starting states to check that we are converging to a global minimum.

\begin{figure}
    \centering
    \includegraphics[width=\columnwidth]{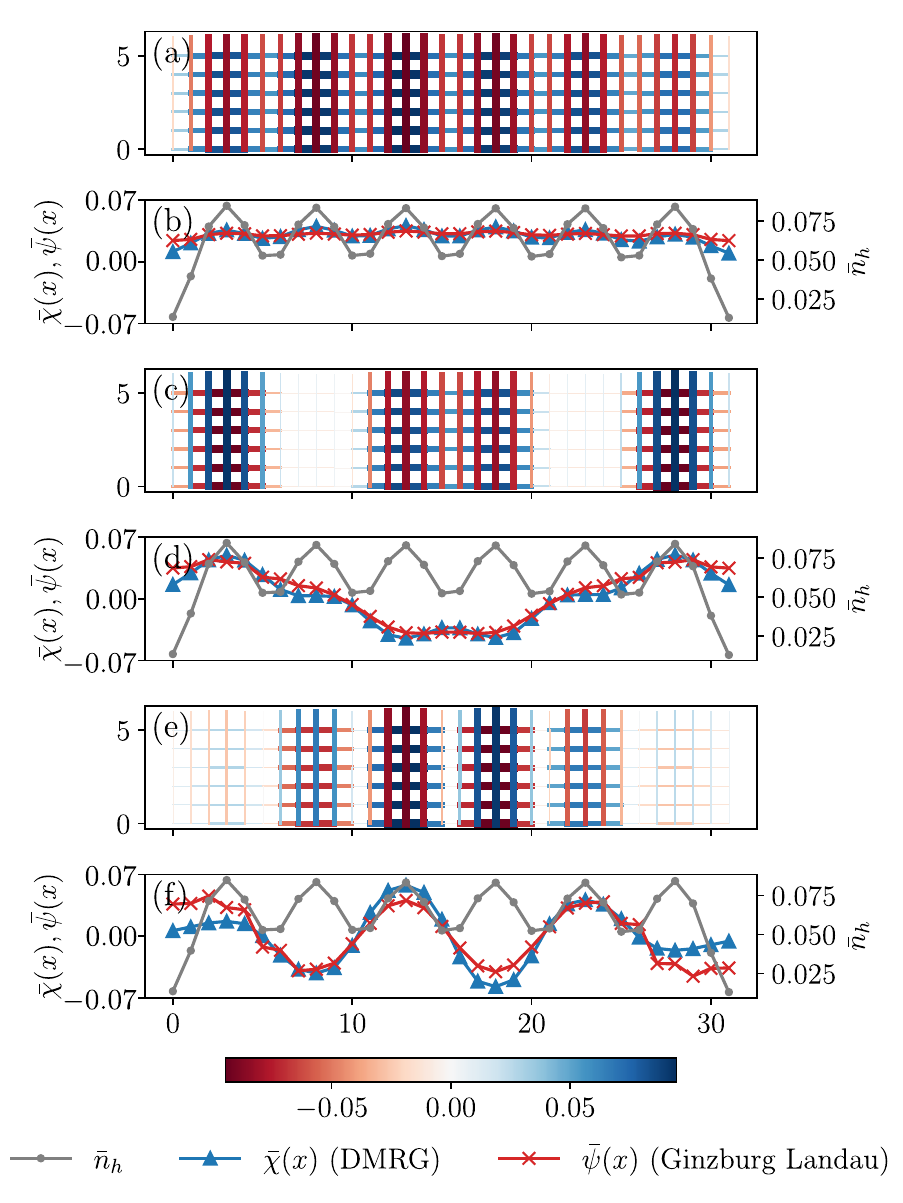}
    \caption{Selected macroscopic wave functions of the Cooper condensate fragments on a $32 \times 6$ cylinder at $t^\prime/t=0.2$, $U/t=10$, and doping $p=1/16$ corresponding to six stripes. (a, c, e) Macroscopic wave functions $\chi_k(\bm{r},\mu)$ as in \cref{eq:nnsingletpairingmatrixdecomp} with index $k=0, 2, 5$. The color and width of the bars denote the value of $\chi_k(\bm{r})$ on the links of the lattice. (b, d, f) $y$-averaged hole-density $\bar{n}_h=(\sum_y n_h(x,y))/W$ of the ground state (grey, dotted) and comparison between the $y$-average of the macroscopic wave functions $\bar{\chi}_k(\bm{r})$ from DMRG and model wave functions $\bar{\psi}_k(\bm{r})$ from the intertwined Ginzburg-Landau theory with $\alpha=-1.3$ and $\beta=13$. We report close agreement between the DMRG data and the solutions from Ginzburg-Landau theory.}
    \label{fig:dmrgresults}
\end{figure}

For the $L=24$ system with open boundary conditions we insert $n_h = 4$ holes in the system, corresponding to a hole-doping $p=1/36$. We start by stabilizing the ground state without any magnetic flux, $\Phi = 0$, at a bond dimension $D=3500$, and then scale up the magnetic flux to $\Phi = 4\pi$ at $D = 4000$, using the previously obtained ground state as the initial state for DMRG. For the $L = 32$ system with cylindrical boundary conditions, we insert $n_h = 12$ holes, corresponding to a hole-doping of $p=1/16$, and we focus on the $\Phi=0$ case only, to better showcase how the different superconducting condensates arrange themselves over the different stripes and demonstrate consistency with the intertwined Ginzburg-Landau results. For this larger system we show results at a bond dimension $D = 8000$.

%In both cases, we run the simulations until we reach a convergence in energy of $\epsilon = 10^{-5}$, with a truncation error in the MPS bond dimension of $\sim 8\cdot 10^{-5}$\aw{not particularly good}. To compute the full two-body density matrix $\rho_S(\bm{r}_i, \mu | \bm{r}_j, \beta)$ we use an optimized MPS algorithm with a reduced scaling of $\mathcal{O}(N^4)$ compared to the naive $\mathcal{O}(N^5)$ and. The details of the numerical procedure are explained in Appendix \ref{num_details}.

We plot the spectrum $\varepsilon_k$ of $\rho_s$  as defined in \cref{eq:nnsingletpairingmatrixdecomp} in Fig.~\ref{fig:spectra} for both parameter sets. In analogy with our previous results on the $t$-$t^\prime$-$J$ model~\cite{Wietek2022}, we observe multiple dominant eigenvalues above a continuum. Again, the number of dominant eigenvalues matches exactly the number of CDW maxima of the ground state. Thus, the system exhibits the presence of $n_h/2$ fragmented condensates, irrespective of the boundary conditions and the presence or absence of a background magnetic field at these parameter values. 

To show the interplay between the CDW and the superconducting order we plot the hole density $n_h(\bm{r})$ superimposed on the macroscopic wave functions. For the $L=24$ system with open boundary conditions, we show the macroscopic eigenvectors $\chi_0(\bm{r}, \mu), \chi_1(\bm{r}, \mu)$ in Fig.~\ref{fig:wfs_hfield}(a),(b), where the angle and the color of the arrows represent the phase of $\chi_k(\bm{r}, \mu)$ and the length of its amplitude respectively. We observe the macroscopic wave functions to be supported on the hole-rich regions of the ground state. Compared to an expected $d$-wave pattern for $\Phi=0$, a finite magnetic field induces rotation in the complex phase that increases by increasing magnetic flux $\Phi$.

For the $L = 32$ system with cylindrical boundary conditions the three macroscopic eigenvectors $\chi_0(\bm{r}, \mu), \chi_2(\bm{r}, \mu), \chi_5(\bm{r}, \mu)$ are shown in Fig.~\ref{fig:dmrgresults}, both in the $x-y$ plane and averaged over the $y$-direction via
\begin{equation}
    \bar{\chi}_k(x) = \frac{1}{4W}\sum_{y=1}^{W} \sum_{\mu=\pm \hat{\bm{x}},\pm \hat{\bm{y}}}
    (-1)^{\mu\cdot\hat{\bm{x}}}\chi_k((x, y), \mu) .
\end{equation}
Since $\chi_k(\bm{r}, \mu)$ is real for $\Phi=0$, we omit the arrows and only employ the color to define the sign. In all cases, the $d$-wave pattern of the pairing and the disposition of the superconducting pairing along the density stripes are clearly visible in alternating red and blue colors for $\mu=\hat{\bm{x}},\hat{\bm{y}}$. In particular, for every different condensate, the macroscopic wave function has a different spatial modulation and correspondingly a different number of nodes.

\begin{figure*}[t]
    \centering
    \includegraphics[width=0.9\textwidth]{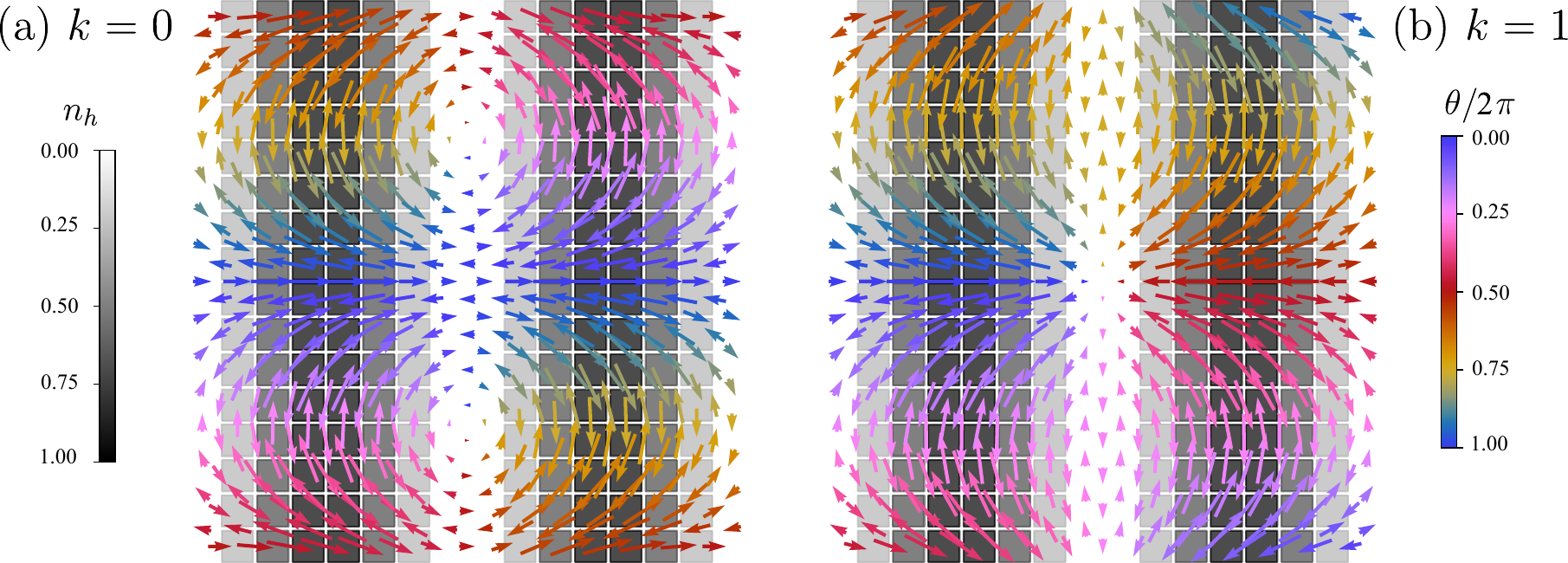}
    \caption{Both solutions $\psi_k(\bm{r}, \mu)$ of the intertwined Ginzburg-Landau equation (\cref{eq:glfunctional}) with $k=0$ (a) and $k=1$ (b). We choose $\alpha=-2.331$ and $\beta=14.465$ on an $L \times L$ lattice with $L=16$ with a total magnetic flux of $\Phi=1.64 \cdot 2\pi$ through the sample and open boundary conditions. The CDW is assumed to be of the form $n_h(\bm{r})=\sin^2(2\pi x / L)$. The grayscale of the background squares is set by the hole density $n_h(\bm{r})$. The color and length denote the phase and amplitude of $\psi_k(\bm{r}, \mu)$, where for $\mu=\hat{\bm{y}}$ we shift the phase by $\pi$ to reflect the $d$-wave nature of the order parameter.  We observe pinned vortices between the stripes in the $k=0$ condensate.}
    \label{fig:full_2d}
\end{figure*}

\section{Intertwined Ginzburg-Landau theory}

In the following, we demonstrate that the macroscopic wave functions $\chi_k(\bm{r}, \mu)$ of the fragmented Cooper condensates are well described by solutions of an intertwined Ginzburg-Landau functional of the form
\begin{equation}
\label{eq:glfunctional}
    \mathcal{F}[\psi] = \alpha(\bm{r})|\psi|^2 + \frac{\beta}{2} |\psi|^4 + \frac{1}{2m^*} \left| \left(-i \hbar \vec{\nabla} + 2e \bm{A}\right) \psi \right|^2.
\end{equation}
The key aspect in \cref{eq:glfunctional} is the position dependence of the mass term $\alpha(\bm{r})$, which is chosen to be
\begin{equation}
    \alpha(\bm{r}) = \alpha n_h(\bm{r}),
\end{equation}
so the charge and phase degrees of freedom are intertwined through directly coupling to the hole density $n_h(\bm{r})$. This action is invariant under a gauge transformation
\begin{equation}
    \psi(\bm{r}) \rightarrow e^{2i\phi(\bm{r})}\psi(\bm{r}), \quad \bm{A}(\bm{r}) \rightarrow \bm{A}(\bm{r}) - \nabla\phi(\bm{r}),
\end{equation}
reflecting the transformation rule of Cooper pairs. Minimizing the functional $\mathcal{F}[\psi]$ by solving
\begin{equation}
\label{eq:functionalderivative}
\frac{\delta \mathcal{F}[\psi]}{\delta \psi}=0,
\end{equation}
yields the time-independent Ginzburg-Landau equations
\begin{equation}
\label{eq:glequation}
    \frac{1}{2m^*}\left( -i \hbar \vec{\nabla} + 2e \bm{A}\right)^2 \psi = \alpha(\bm{r}) \psi + \beta |\psi|^2 \psi.
\end{equation}
Importantly, the criterion \cref{eq:functionalderivative} from which \cref{eq:glequation} is derived is valid for every \textit{local} minimum of the functional \cref{eq:glfunctional}. Hence, whenever there are multiple solutions to \cref{eq:glequation} there are also multiple local minima of \cref{eq:glfunctional}.
In the absence of both the nonlinear term ($\beta=0$) and the vector potential ($\bm{A}=0$), the Ginzburg-Landau equation (\cref{eq:glequation}) reduces to the linear time-independent Schr\"odinger equation. For periodic boundary conditions, in the presence of a periodic modulation of the mass term
\begin{equation}
\alpha(\bm{r} + \bm{\lambda}) = \alpha(\bm{r}),
\end{equation}
Bloch's theorem~\cite{Bloch1929} states that this equation has multiple solutions labeled by the wave number $\bm{k}$,
\begin{equation}
    \psi_{\bm{k}}(\bm{r}) = e^{i\bm{k}\cdot\bm{r}}u(\bm{r}),
\end{equation}
where $u(\bm{r} + \bm{\lambda}) = u(\bm{r})$. 
%For example, when choosing a periodic potential of the form,
%\begin{equation}
%    \alpha(x) = a + b\cos(x),
%\end{equation}
%in one dimension this yields the Mathieu differential equation with well-known analytical solutions~\cite{NIST:DLMF}. 
Naturally, the question arises whether an analog of Bloch's theorem still holds in the presence of a nonlinearity ($\beta \neq 0$). This has already been addressed in the literature, where generalizations of Bloch's theorem with additional nonlinearities have been proven~\cite{Serkin2001,Haus1999,Louis2003}. Note that nonlinear Schr\"odinger equations in periodic potentials naturally occur when describing Bose-Einstein condensates in optical lattices, see e.g.~\cite{Choi1999,Louis2003,Denschlag2002,Bloch2005}. 

In the presence of both a nonlinearity and the magnetic vector potential in two dimensions, we solve \cref{eq:glequation} numerically. Our approach is to solve the equivalent problem of minimizing $\mathcal{F}[\psi]$ via a local minimization algorithm and verifying that the functional derivative in \cref{eq:functionalderivative} vanishes. Different solutions corresponding to distinct Bloch waves can be obtained by starting the optimization procedure with distinct initial configurations within the basin of attraction of the particular solution, cf. \cref{app:gldetails}. We employ the BFGS algorithm~\cite{Broyden1970,Fletcher1970,Goldfarb1970,Shanno1970} for local optimization and verify the minimization condition \cref{eq:functionalderivative} by monitoring the norm of the Jacobian matrix, which we require to attain a numerical value $<10^{-7}$ in all studied geometries. 

To compare the solutions of the intertwined Ginzburg-Landau equation (\cref{eq:glequation}) to the macroscopic wave functions from DMRG, we solve for $\psi_k(\bm{r})$ on the vertices of the square lattice. We then define the $d$-wave order parameter ansatz on the links of the lattice as
\begin{equation}
\label{eq:ginzburg_dwave}
    \psi_k(\bm{r}, \mu) = 
    \begin{cases}
    +\left[\psi_k(\bm{r}) + \psi_k(\bm{r} + \mu)\right]/2
    \quad \text{if} \quad \mu = \hat{\bm{x}} ,\\ 
    -\left[\psi_k(\bm{r})  + \psi_k(\bm{r} + \mu)\right]/2
    \quad \text{if} \quad \mu = \hat{\bm{y}}. 
    \end{cases}
\end{equation}

With this definition, the $d$-wave nature of the order parameter $\psi_k(\bm{r}, \mu)$ is predetermined. We compare the intertwined Ginzburg-Landau wave functions with the numerical results in Fig.~\ref{fig:wfs_hfield}(b) for a finite magnetic flux $\phi$. At this set of parameters, DMRG yields two fragments of the condensate while the Ginzburg-Landau equation exhibits two distinct solutions, which we find using different initial states when minimizing the Ginzburg-Landau functional, cf. \cref{fig:wfs_convergence}. Thus, there is a one-to-one correspondence between the superconducting fragments and the distinct solutions of the Ginzburg-Landau theory. Moreover, we observe that both the periodicity in $\chi_0(\bm{r}, \mu)$ and $\chi_1(\bm{r}, \mu)$ and the rotation due to the magnetic field are correctly captured by the Ginzburg-Landau solutions $\psi_0(\bm{r}, \mu)$ and $\psi_1(\bm{r}, \mu)$ and observe accurate agreement in \cref{fig:wfs_hfield}. To obtain such an agreement, the two model parameters $\alpha$ and $\beta$ have to be optimized. We performed a full parameter scan to determine optimal values of $\alpha=-2.331$, $\beta=14.465$, cf. \cref{fig:parameter_fits}.

Next, we consider the case of multiple stripes without a magnetic field ($\Phi=0$) on a $32 \times 6$ cylinder. As we only obtain DMRG results for open boundary conditions in the $x$-direction, the corresponding Ginzburg-Landau equation coupling to the hole density from DMRG is not translationally invariant. Thus, we do not immediately expect Bloch waves as solutions. Instead, we find solutions that are localized on the individual stripes and compute their Fourier transform at a momentum $k$ of the superlattice set by the charge density wave which serves as ansatz wave functions $\psi_k(\bm{r}, \mu)$ to compare to DMRG. A detailed discussion of this construction is given in \cref{app:glbloch}. \cref{fig:dmrgresults} again shows a close agreement between the ansatz wave functions $\psi_k(\bm{r}, \mu)$ from the intertwined Ginzburg-Landau theory and the condensate fragments $\chi_k(\bm{r}, \mu)$. Moreover, this comparison reveals that the different fragments of the condensate can be labeled by the quasi-momentum of the wave function on the superlattice given by the charge density wave. This is the reason we used the label `$k$' for both enumerating the fragments $\chi_k(\bm{r}, \mu)$ and model wave functions $\psi_k(\bm{r}, \mu)$. In fact, we observe a `dispersion', in the sense that the smaller values of $k$ have a larger condensate fraction, i.e. $k^\prime < k$ implies $\varepsilon_{k^\prime} > \varepsilon_k$, reflecting the fact that the uniform condensate is the most dominant condensate.

The intertwined Ginzburg-Landau theory allows us to make predictions for larger systems not accessible by DMRG simulations. As an example, we report in Fig.~\ref{fig:full_2d} the solutions of \cref{eq:glfunctional} on a $16\times16$ grid with a CDW of the form $n_h(\bm{r})=\sin^2(2\pi x / L)$. For this larger size we can see how, on the $k=0$ condensate, vortices are effectively pinned between the stripes for the chosen set of parameters $\alpha=-2.331$ and $\beta=14.465$. This set of parameters is the optimal choice when fitting to the DMRG results in \cref{fig:wfs_hfield}. However, we expect the behavior of the intertwined Ginzburg-Landau equation to be more complex in general. Besides the conventional coherence length and penetrations depth, another natural length scale that can be defined is the period of the charge modulation. Moreover, we also expect the depth of the potential to play an important role. A more detailed study of the behavior when varying these scales will be the subject of future investigations. 

\section{Discussion and Conclusions}
The coexistence of CDWs and superconductivity has by now been reported by numerous numerical studies of doped Hubbard- and $t$-$J$-like models~\cite{Gong2021,Jiang2021,HCJiang2021,Jiang2023,Jiang2023b,Xu2023,Lu2023,Wang2022,Ponsioen2023}. By employing the Penrose-Onsager criterion for superconductivity, one of the authors previously reported that this coexistence yields a fragmentation of the superconducting condensate in the case of the $t$-$t^\prime$-$J$ model on the square lattice~\cite{Wietek2022}. This is seen from investigating the eigenvalues and eigenvectors of a suitably chosen two-body density matrix, which correspond to the condensate fractions and the macroscopic wave functions of the Cooper pairs, where more than one dominant eigenvalue is observed. In the present manuscript, we corroborated these findings by demonstrating that fragmentation is not just a peculiarity of the $t$-$J$ model but is observed in the phase diagram $t$-$t^\prime$-$U$ Hubbard model. More importantly, we propose a macroscopic field theory that describes the wave functions of the fragmented condensate with remarkable precision. The key proposition is the intertwined Ginzburg-Landau theory \cref{eq:glfunctional}, where the mass term of the superconducting order parameter is coupled to a static field proportional to the hole density of the system. The CDW is seen as a periodic background potential that leads to the existence of multiple solutions of the Ginzburg-Landau theory, which correspond to Bloch waves at different momenta. These distinct solutions of the intertwined Ginzburg-Landau theory have been shown to precisely describe the individual fragments of the superconducting condensates, cf. \cref{fig:wfs_hfield,fig:dmrgresults}. We would like to emphasize that the model we propose only features two parameters $\alpha$ and $\beta$ (up to an overall constant factor $m^*$) and can therefore be considered a minimal model coupling the charge density to the superconducting order. 

Our findings give rise to an intuitive picture of the behavior of pairs in the stripe-fragmented superconductor. Here, a Cooper pair is in an unequal-weight superposition of the condensate wave functions at different momenta. The individual macroscopic wave functions constitute \textit{local} minima of the Ginzburg-Landau free energy functional \cref{eq:glfunctional} and do not necessarily yield the same value of the Ginzburg-Landau free energy. Nevertheless, the Cooper pair can ``tunnel'' between these different minima. An energy difference between these minima is then reflected by a different occupation number in these modes, just as we observe in our DMRG data.  We numerically verified that the uniform condensate solution is a \textit{global} minimum of the Ginzburg-Landau functional, which explains why this mode is populated with the highest condensate fraction, whereas finite momentum solutions have higher values of the Ginzburg-Landau free energy and thus a smaller condensate fraction. 

A coupling between a CDW and the superconducting order parameter has been suggested previously~\cite{Cai2017,Gabovich2009,Balseiro1979,Lian2018,Chang2012,Castellani1996}. Importantly, novel pump-probe experiments allow for a precise measurement of this coupling employing Fano interference~\cite{Chu2020,Chu2023}, and a coupling between the CDW and the superconducting order has recently been measured in \ch{NbSe2}~\cite{Feng2022} and the cuprate superconductors~\cite{Chu2020,Chu2023}. Thus, there is strong evidence that coupling between CDWs and superconducting order parameters is realized in actual strongly correlated superconductors. 

The final question we would like to address is whether a fragmented Cooper condensate could be observed in materials. While our particular choice of parameters in the $t$-$t^\prime$-$U$ might not be the set of couplings expected in e.g. the cuprate superconductors, it is very likely that such supersolid phases with fragmented condensates can occur in more complex scenarios, which then capture the physics of actual materials more accurately. Several studies employing scanning Josephson tunneling spectroscopy have recently reported the emergence of a pair-density wave state~\cite{Agterberg2020} in \ch{Bi2Sr2CaCu2O}$_{8+x}$~\cite{Hamidan2016,Edkins2019}, in transition metal dichalcogenides~\cite{Liu2021}, or the heavy-fermion superconductor \ch{UTe2}~\cite{Gu2023}. A peak in the superconducting order parameter at finite momentum has been detected, indicating the emergence of a pair density wave. Besides this finite momentum peak, however, a zero momentum contribution was also measured. We would like to point out that this scenario bears a strong resemblance to our findings, as we propose that the Cooper pairs are in a superposition of condensates with different momentum quantum numbers. Hence, if a fragmented superconductor were to be realized, further peaks at different momenta could indicate this state. Moreover, a prediction of our intertwined Ginzburg-Landau theory would be that the superconducting vortices in a magnetic field are pinned between the CDWs, as shown in \cref{fig:full_2d}, which we propose as a potential hallmark experimental signature of this phase. 

\begin{acknowledgements}
The DMRG simulations have been performed using the Julia~\cite{bezanson2017julia} version of the ITensor library~\cite{itensor,itensor-r0.3} and calculations of the singlet-pairing density matrix were performed with the Julia library \emph{ITensorCorrelators.jl}~\cite{ITensorCorrelators}. The code to solve the intertwined Ginzburg-Landau equations is accessible via its GitHub repository~\cite{GLPack}. We thank Stefan Kaiser, Matthias Vojta, J\"org Schmalian, Andrew Millis, Antoine Georges, Roderich Moessner, and Christina Kurzthaler for stimulating discussions. The Flatiron Institute is a division of the Simons Foundation. A.W. acknowledges support by the DFG through the Emmy Noether program (Grant No.\ 509755282). N.B. acknowledges support from a ``la Caixa'' Foundation fellowship (ID 100010434, code LCF/BQ/DI20/11780033). ICFO group acknowledges support from: ERC AdG NOQIA; MICIN/AEI (PGC2018-0910.13039/501100011033, CEX2019-000910-S/10.13039/501100011033, Plan National FIDEUA PID2019-106901GB-I00, FPI; MICIIN with funding from European Union NextGenerationEU (PRTR-C17.I1): QUANTERA MAQS PCI2019-111828-2); MCIN/AEI/ 10.13039/501100011033 and by the ``European Union NextGeneration EU/PRTR'' QUANTERA DYNAMITE PCI2022-132919 within the QuantERA II Programme that has received funding from the European Union’s Horizon 2020 research and innovation programme under Grant Agreement No 101017733 Proyectos de I+D+I ``Retos Colaboración'' QUSPIN RTC2019-007196-7); Fundaci\'o Cellex; Fundaci\'o Mir-Puig; Generalitat de Catalunya (European Social Fund FEDER and CERCA program, AGAUR Grant No. 2021 SGR 01452, QuantumCAT \ U16-011424, co-funded by ERDF Operational Program of Catalonia 2014-2020); Barcelona Supercomputing Center MareNostrum (FI-2023-1-0013); EU (PASQuanS2.1, 101113690); EU Horizon 2020 FET-OPEN OPTOlogic (Grant No 899794); EU Horizon Europe Program (Grant Agreement 101080086 — NeQST), National Science Centre, Poland (Symfonia Grant No. 2016/20/W/ST4/00314); ICFO Internal ``QuantumGaudi'' project; European Union's Horizon 2020 research and innovation program under the Marie-Skłodowska-Curie grant agreement No 101029393 (STREDCH) and No 847648 (``La Caixa” Junior Leaders fellowships ID100010434: LCF/BQ/PI19/11690013, LCF/BQ/PI20/11760031, LCF/BQ/PR20/11770012, LCF/BQ/PR21/11840013). Views and opinions expressed are, however, those of the author(s) only and do not necessarily reflect those of the European Union, European Commission, European Climate, Infrastructure and Environment Executive Agency (CINEA), or any other granting authority. Neither the European Union nor any granting authority can be held responsible for them.

\end{acknowledgements}

\appendix

\section{Details on DMRG simulations} \label{num_details}

\begin{figure}
    \centering
    \includegraphics[width=0.9\columnwidth]{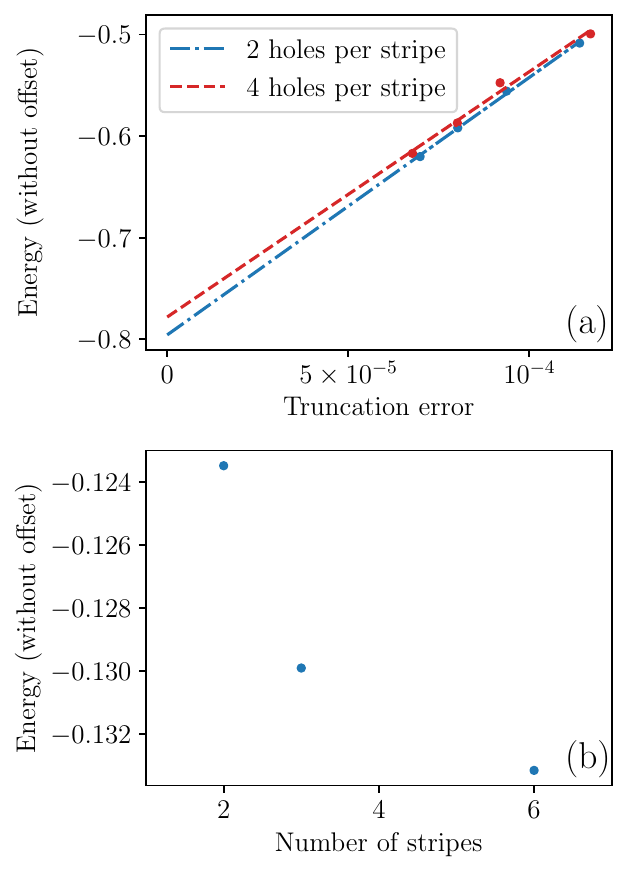}
    \caption{(a) Scaling of the energy with the truncation error for two different ansatzes in a $24\times6$ system with four holes. (b) Energy behaviour for different ansatzes at fixed bond dimension $\chi =4000$ for a $32\times6$ system with 12 holes}
    \label{fig:scaling}
\end{figure}

\begin{figure*}
\includegraphics[width=\textwidth]{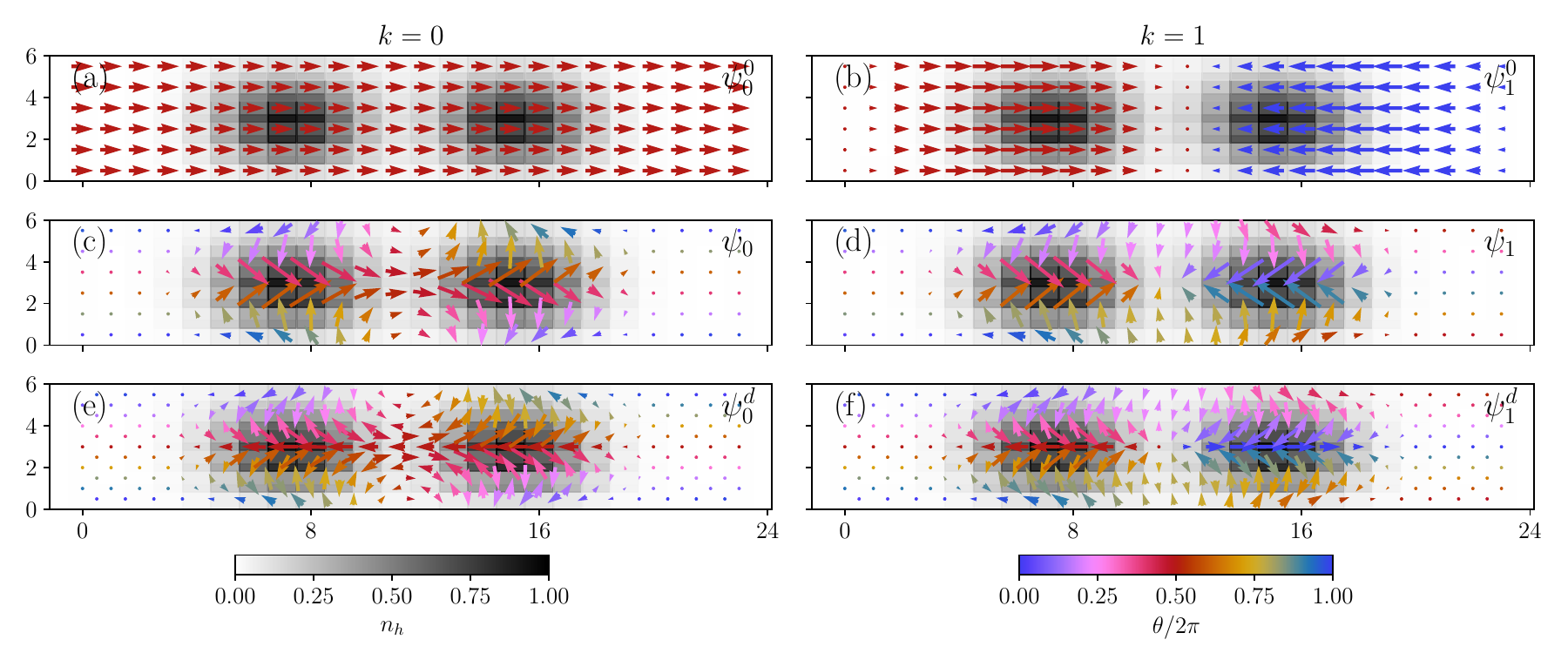}
\caption{(a,b) Initial ansatz wave functions $\psi^0_k(\bm{r})$ used as a starting point in the minimization of the Ginzburg-Landau functional \cref{eq:glfunctional}. (c,d)
Obtained local minima $\psi_k(\bm{r})$ from these initial states after optimization. (e,f) Resulting $d$-wave order parameters $\psi_k(\bm{r}, \mu)$ as in \cref{eq:ginzburg_dwave} defined on the links of the lattice.}
\label{fig:wfs_convergence}
\end{figure*}

To ensure consistent convergence of DMRG for the systems we have studied, some care must be taken in choosing the initial states of the variational optimization. Here we describe our strategy for obtaining initial states. For systems of width $W=4$, we found that DMRG is not sensitive to the initial starting states and we use random initial states. Independent of the starting state, we find that the ground state is antiferromagnetic with a checkerboard spin pattern, and the holes bind in pairs and distribute evenly along the width of the system.

For width $W = 6$, we use starting states that are inspired by the ground states we find with DMRG at $W = 4$, which we find provide better convergence than random initial states. We have tried different starting states to test that our choice of starting state doesn't lead to a biased result. As an example of this, in Fig.~\ref{fig:scaling}(a), we show results from DMRG calculations of an $L=24$ system with open boundary conditions and an equal number of total holes which were biased by the starting states to have different hole configurations (two holes per stripe with more stripes or four holes per stripe fewer stripes). By extrapolating the energy as a function of DMRG truncation error~\cite{Hubig2018}, we can conclude that the state with 2 holes per stripe is lower in energy.

In Fig.~\ref{fig:scaling}(b), for an $L=32$ system with cylindrical boundary conditions and $n_h = 12$ holes, we compare the variational minimum energy for three different hole configurations (two, four, and six holes per stripe) at fixed bond dimension and as before we see that the energy is lowest for the state with two holes per stripe. Throughout this work, we therefore show results using states that have two holes per stripe.

A finite magnetic field $\phi$ in the Hamiltonian Eq.~\ref{hubbard_ham} introduces complex coefficients. In DMRG, this leads to tensors that have complex elements, and therefore the computation time increases by a factor of approximately four compared to the case of zero magnetic field ($\phi=0$) where tensors with real elements can be used. To improve the convergence time of calculations with finite magnetic field, we first compute the ground state at some fixed bond dimension without a magnetic field and use the state found by DMRG as a starting state for calculations at finite magnetic field.

Care must be taken to efficiently compute the two-body singlet-pairing density matrix $\rho_S$. Naively, without the use of caching and sparsity, computing every element of a general two-body density matrix on a system of size $N = \mathcal{O}(L)$ (where for simplicity we assume a fixed width $W$) would scale as $\mathcal{O}(N^5)$ since there are $\mathcal{O}(N^4)$ elements and computing a single element requires $\mathcal{O}(N)$ tensor contractions, which would quickly become impractical. Luckily, there are multiple ways we can reduce this scaling. First of all, we only need to compute the pairing on neighboring sites, which reduces the scaling to $\mathcal{O}(N^3)$ if no caching is used. If caching of intermediate tensor contractions is used across the computation of different elements, this scaling can be further reduced to $\mathcal{O}(N^2)$. The code we use, which can compute general $n$-body correlators and automatically cache intermediate tensors involved in computing different elements, can be found at~\cite{ITensorCorrelators}.

\subsection{Scaling of density wave pattern with system size}

\begin{figure*}
    \centering
    \includegraphics[width=\textwidth]{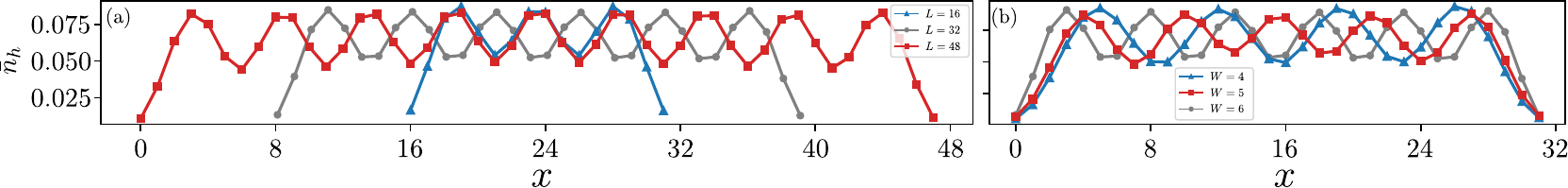}
    \caption{Density wave pattern for different system sizes at fixed hole density $p=1/16$ and $\Phi=0$. The grey dots represent the $32\times6$ system studied in the main text. (a) By increasing the length of the cylinder to $L=48$ (red squares) or decreasing it to $L=16$ (blue triangles), we do not see a sizable change of the hole-density periodicity or amplitude, excluding a boundary effect nature. (b) By decreasing the width of the cylinder to $W=5$ (red squares) or $W=4$ (blue triangles), the amplitude of the density modulations does not change.}

    %The different periodicity is due to the different amount of holes present in the system when fixing the filling. In both cases, the density pattern is compatible with a density wave with a number of stripes equal to $n_h/2$ where $n_h$ is the number of holes.
    
    \label{fig:scalings_density}
\end{figure*}

To certify that the density pattern we found is indeed due to the presence of a CDW coexisting with superconductivity, and to exclude that it is caused by Friedel oscillations near the boundaries of the system, we performed a scaling of the hole density in both the longitudinal and transverse directions of our system. We focused on a system at fixed hole-doping $p=1/16$, $\Phi=0$, $U/t=10$, $t'/t=0.2$, with periodic boundary conditions on the $y$-axis, taking as a reference the results for the $32\times 6$ cylinder, as pictured in Fig. \ref{fig:dmrgresults} in the main text. 

By keeping $W=6$ and increasing $L$ to $48$, or reducing it to $16$, we see that the density pattern is compatible with a CDW with a number of stripes equal to $n_h/2$ where $n_h$ is the number of holes. Most importantly, the periodicity and the amplitude of the hole-density modulation does not change when changing $L$, excluding a finite size nature of the oscillations. This is shown in Fig. \ref{fig:scaling}(a). In this scenario, we performed simulations for bond dimensions up to $\chi=7000$, corresponding to a truncation error $\varepsilon\sim 5\times 10^{-5}$.

Analogously, keeping $L=32$ and decreasing $W$ to $5$ and $4$ does not change qualitatively the behavior of the density wave, as shown in Fig.~\ref{fig:scaling}(b). In this case, the periodicity of the density wave changes as a bigger number of holes can be accommodated in the system at fixed density, when increasing $W$.

\subsection{Results for $t'/t-0.2$}

\begin{figure}
    \centering
    \includegraphics[width=0.7\columnwidth]{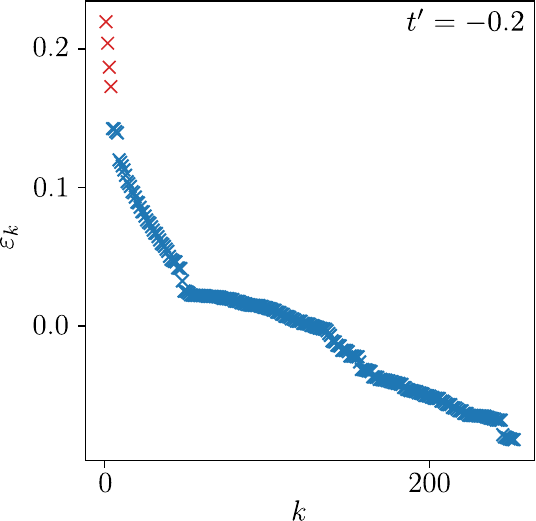}
    \caption{Fragmentation of the Cooper condensate from the spectrum of the two-body density matrix on the ground state of a $32\times4$ cylinder with negative $t^\prime/t=-0.2$, doping $p=1/16$ corresponding to $n_h=8$, and $U/t = 10$. The dominant eigenvalues, indicated by red crosses, are less prominent when compared to the $t'/t=0.2$ case.}
    \label{fig:fragment_negativetp}
\end{figure}

\begin{figure*}
    \centering
    \includegraphics[width=0.8\linewidth]{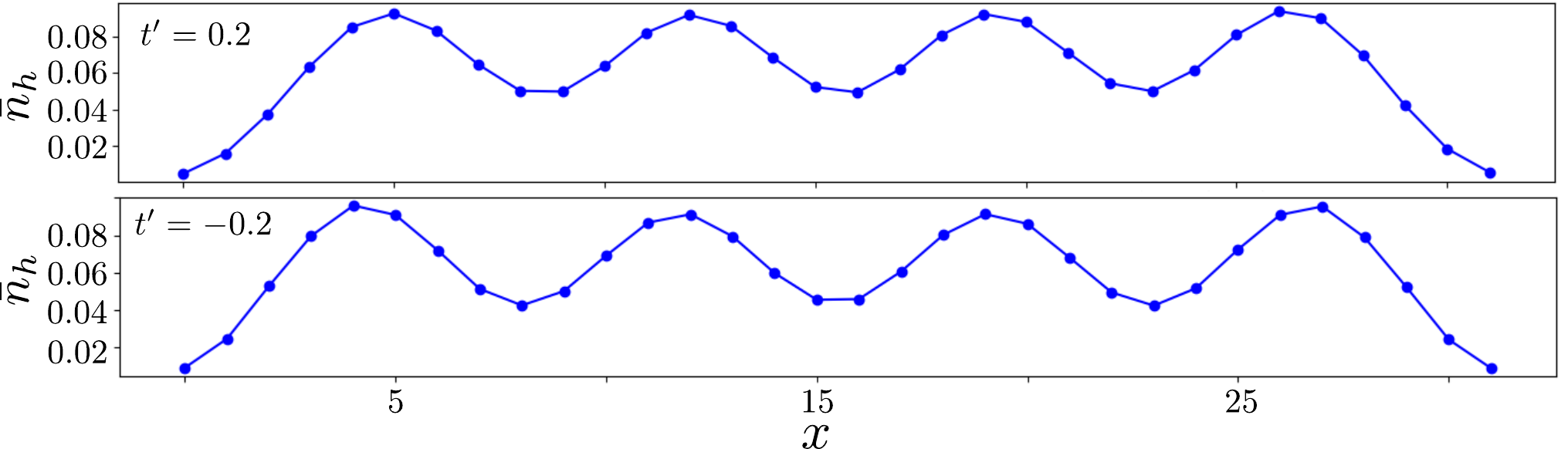}
    \caption{Comparison of the density wave pattern of a $32\times 4$ system on a cylinder with $t^\prime/t=0.2$ or $t'/t=-0.2$ for $U/t=10$. In both cases the system shows the same number of stripes with comparable hole density modulations.}
    \label{fig:density_compar}
\end{figure*}

In the main text, we focused on a value of the next-nearest-neighbor hopping $t'/t=0.2>0$. However, it is interesting to understand what happens for $t'<0$, as this scenario is relevant for the hole-doped cuprates \cite{Xu2023,SilvaNeto2016}. In this case, we find that there is still a fragmentation of the Cooper pair condensate, as shown in Fig.~\ref{fig:fragment_negativetp} for a $32\times 4$ cylinder, with $t'/t-0.2$, at doping $p=1/16$, corresponding to $n_h=8$. Compared to the $t'/t=0.2$ case, the fragmentation is less pronounced, as the dominant eigenvalues are less prominent. Just as for $t'>0$, the number of dominant eigenvalues correspond to the number of CDW peaks in the ground state, as seen in Fig.~\ref{fig:density_compar}. We also note that the modulation in hole density is very similar to the case of $t'>0$. 

\section{Details on solving the Ginzburg-Landau equations}
\label{app:gldetails}

We describe how to numerically attain the distinct solutions corresponding to the distinct Bloch wave instanton solutions of \cref{eq:glequation} in the main text. These are computed by numerically minimizing the parent Ginzburg-Landau functional \cref{eq:glfunctional} with different initial starting vectors. The procedure is illustrated in \cref{fig:wfs_convergence}. To obtain a uniform solution we start with a constant initial state $\psi_0^0$, e.g.
\begin{equation}
    \psi_0^0(\bm{r}) = 1.0,
\end{equation}
as shown in \cref{fig:wfs_convergence}(a). $\psi(\bm{r})$ is a function with complex values on the vertices of the lattice. The values of hole-density $n_h(\bm{r})$ on the lattice are taken from measurements of the ground state from DMRG. We obtain a local minimum of the Ginzburg-Landau functional \cref{eq:glfunctional} by employing the BFGS algorithm~\cite{Broyden1970,Fletcher1970,Goldfarb1970,Shanno1970} as implemented in SciPy~\cite{Scipy}. Convergence is tested by postulating the norm of the Jacobian matrix to be of size $<10^{-7}$. This allows us to obtain the wave function $\psi_0(\bm{r})$, as shown in \cref{fig:wfs_convergence}(b). To compare this to the macroscopic wave function from DMRG, we compute the corresponding $d$-wave order parameter on the links of the lattice by \cref{eq:ginzburg_dwave}. The final model wave function $\psi_0(\bm{r}, \mu)$ is then shown in \cref{fig:wfs_convergence}(c).

To compute different solutions we initialize the minimization procedure with a starting state of the form
\begin{equation}
    \psi_k^0(x, y) = \sin(2\pi k x / L ).
\end{equation}
The solution shown in \cref{fig:wfs_hfield}(d) of the main text is obtained when setting $k=1$. The corresponding initial configuration is shown in \cref{fig:wfs_convergence}(d). After finding the local minimum associated with this initial solution, we obtain  $\psi_1(\bm{r})$ as shown in \cref{fig:wfs_convergence}(e), and finally the $d$-wave order parameter $\psi_1^d(\bm{r})$ shown in \cref{fig:wfs_convergence}(f).

\section{Choice of model parameters $\alpha$ and $\beta$}

\begin{figure}
    \centering
    \includegraphics[width=\columnwidth]{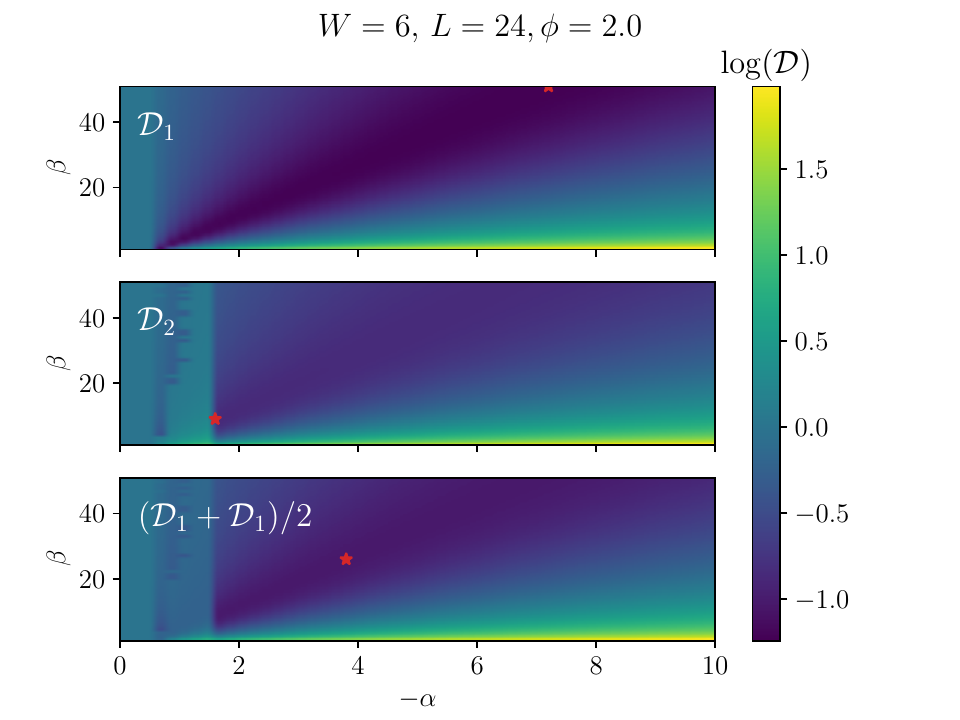}
    \caption{Full parameter scan for the metric \cref{eq:metric} for model parameters $\alpha$ and $\beta$ of the intertwined Ginzburg-Landau equation when comparing to the DMRG ground state shown in \cref{fig:wfs_hfield} of the main text. The red star indicates the minimizing model parameters.}
    \label{fig:parameter_fits}
\end{figure}

Our proposed intertwined Ginzburg-Landau functional in \cref{eq:glfunctional} in the main text only has two parameters $\alpha$ and $\beta$ up to an overall constant factor set by $m^*$. To obtain good agreement we have investigated the dependence of the solutions on these parameters. These have then been compared to the macroscopic wave functions $\chi_k(\bm{r}_i, \mu)$ obtained from DMRG. To find the optimal parameters for $\alpha$ and $\beta$ we consider the following metric comparing both these wave functions:
\begin{equation}
\label{eq:metric}
    \mathcal{D}_k = \min_\theta 
    \parallel 
    \chi_k(\bm{r}_i, \mu) - e^{i\theta} \psi_k(\bm{r}_i, \mu) 
    \parallel ^2.
\end{equation}
The phase $\theta$ is introduced due to the freedom of phase in the solution of both the eigenvalue problem to compute $\chi_k(\bm{r}_i, \mu)$ as well as the U($1$) symmetry of the Ginzburg-Landau functional \cref{eq:glfunctional}. Since only two parameters are involved we can perform a full parameter scan to obtain an optimal fit. The metrics \cref{eq:metric} when fitting the DMRG ground state with $\Phi=4\pi$ on the $24 \times 6$ open boundary sample as in \cref{fig:wfs_hfield} of the main text are shown in \cref{fig:parameter_fits}. The combined metric $\mathcal{D}_0 + \mathcal{D}_1$ is minimized for the parameters $\alpha=-2.331$ and $\beta=14.465$.

\section{Bloch waves from solutions of the Ginzburg-Landau equation in the multiple stripe case}
\label{app:glbloch}

\begin{figure}
    \centering
\includegraphics[width=\columnwidth]{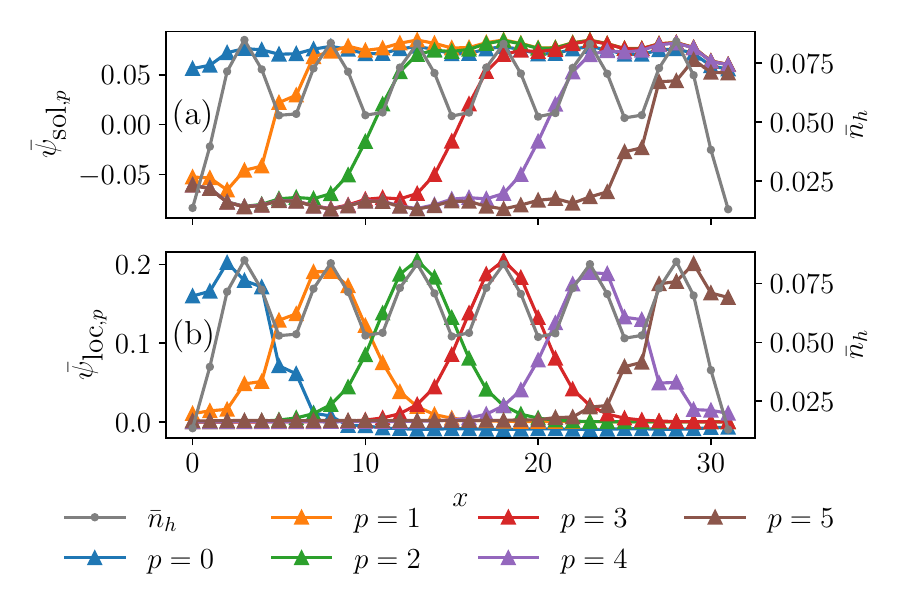}
    \caption{(a) Solutions of the intertwined Ginzburg-Landau equation in the presence of six stripes on a $32 \times 6$ cylinder with open boundary conditions in the $x$-direction. We show the $y$-averaged values of $\psi_k(\bm{r}, \mu)$. For every neighboring pair of stripes a solution exhibiting a ``kink'' at their interface is found. (b) Corresponding localized wave functions $\psi_{\textrm{loc},p}$ as defined in \cref{eq:locwfs}.}
    \label{fig:wf6_construction}
\end{figure}

\begin{figure*}
    \centering
\includegraphics[width=\textwidth]{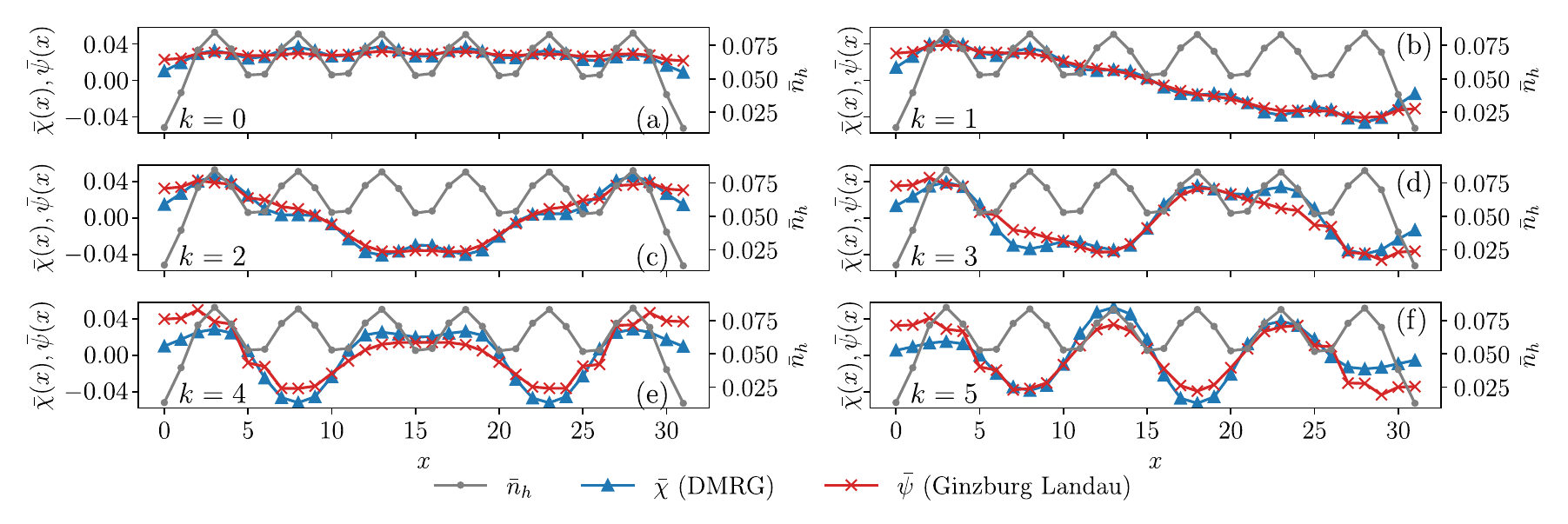}
    \caption{Complete set of macroscopic wave functions of the ground state of $32\times 6$ cylinder at $t^\prime/t=0.2$, $U/t=10$, and doping $p=1/16$, corresponding to six stripes. Blue lines (triangles) denote wave functions from DMRG of the six fragments of the condensate. Red lines are the corresponding solutions $\psi_k(\bm{r})$ of the intertwined Ginzburg-Landau theory as specified in \cref{eq:glblochstates} for $\alpha =-1.3$ and $\beta=13$. Different panels show show the results for different values of the momentum $k$. We report close agreement between the DMRG data and the Ginzburg-Landau ansatz for every single fragment. }
    \label{fig:wf6_momentum}
\end{figure*}

In the case of two stripes in the system, we demonstrated in \cref{fig:wfs_hfield} that the macroscopic wave functions $\chi_k(\bm{r}, \mu)$ of the superconducting condensates in \cref{eq:nnsingletpairingmatrixdecomp} are well described by solutions $\psi_k(\bm{r}, \mu)$ of the intertwined Ginzburg-Landau equations \cref{eq:glequation} with two different initial ansatz wave functions, cf. \cref{fig:wfs_convergence}. For more than two stripes the situation is slightly complicated by the fact that our DMRG simulations are performed on open boundary conditions in the $x$-direction. Hence, the solutions to the Ginzburg-Landau equations do not a priori satisfy Bloch's theorem and, hence, momentum is not a good quantum number. We solve the intertwined Ginzburg-Landau equation (\cref{eq:glequation}) with the hole density given by the ground state from DMRG on the cylindrical boundary conditions. Thereby, we obtain solutions that are either uniform $\psi_{\textrm{sol}, 0}(\bm{r})$ or have a ``kink'' between stripe $i$ and $i+1$, which we denote by $\psi_{\textrm{sol}, p}(\bm{r})$, where $p=1, \ldots, N_{\textrm{str}}-1$ and  $N_{\textrm{str}}$ denotes the number of stripes in the system. Here, $p$ labels the position of the stripe in the $x$ direction. These direct solutions $\psi_{\textrm{sol}, p}(\bm{r})$ in the case of the $32\times 6$ cylinder, with $t^\prime/t=0.2$ and $U/t=10$, are shown in \cref{fig:wf6_construction}(a). In this case, we consider the case of zero magnetic flux, i.e. $\Phi=0$. On an infinitely long cylinder with a perfectly periodic modulation of the charge density, the position of the kink does not impact the value of the free energy functional \cref{eq:glfunctional} due to translational invariance. Hence, all solutions with only one kink are degenerate and the system can choose to form superpositions of these states. We consider the wave functions
\begin{equation}
\label{eq:locwfs}
    \psi_{\textrm{loc},p}(\bm{r}) = \psi_{\textrm{sol},p+1}(\bm{r}) - \psi_{\textrm{sol},p}(\bm{r}),
\end{equation}
which are localized on stripe $p$, as shown in \cref{fig:wf6_construction}(b). In order to construct Bloch wave functions with a given quasi-momentum $k$ we construct the wave functions
\begin{equation}
\label{eq:glblochstates}
    \psi_{k}(\bm{r}) = \sum_{p=0}^{N_{\textrm{str}}-1} \cos(\pi\; p\cdot k / (N_{\textrm{str}}-1)) \psi_{\textrm{loc}, p}(\bm{r}).
\end{equation}
This procedure yields a wave function for $k=0, \ldots, N_{\textrm{str}}-1$, where $k=0$ corresponds to the uniform solution. The $y$-averaged solutions $\bar{\psi}_{k}(x)$ in  case of six stripes for the $32 \times 6$ cylinder is shown for select values of $k$ in \cref{fig:dmrgresults} of the main text and all values of $k$ in \cref{fig:wf6_momentum}. We report close agreement between the Ginzburg-Landau ansatz wave functions $\psi_{k}(\bm{r})$ and the macroscopic wave functions of the dominant condensate fragments from ground state DMRG. Remarkably, the labeling of Bloch states in \cref{eq:glblochstates} exactly corresponds to the ordering of the fragments by their condensate fraction, i.e. the $k=0$ is the most dominant eigenvalue, $k=1$ the second most dominant, etc., until $k=5$ corresponds to the least dominant eigenvalue. 

\section{Two-body reduced density matrix}

\begin{figure}
    \centering
\includegraphics[width=0.8\columnwidth]{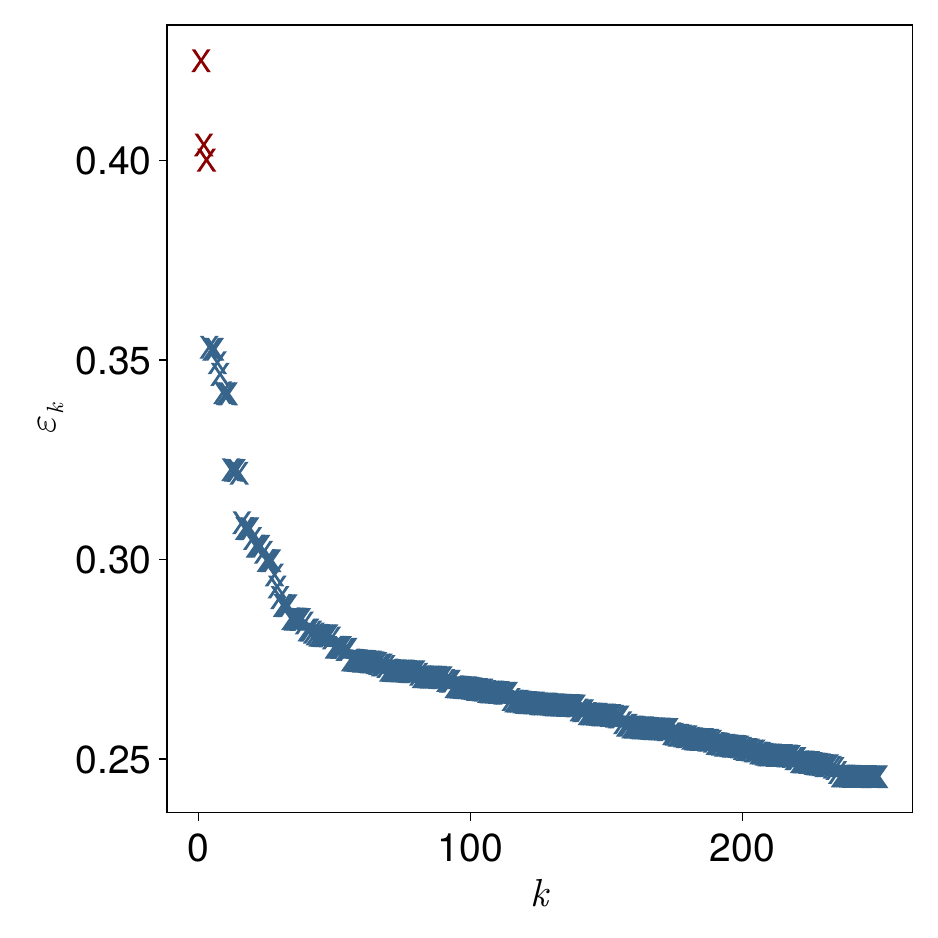}
    \caption{The spectrum of the full 2RDM for a $16 \times 6$ cylinder, at $t'/t=0.2$, $\Phi=0$, and hole-doping $p = 1/16$. For the sake of visualization, only the first $250$ eigenvalues are shown. The dominant eigenvalues are indicated by red crosses.}
    \label{fig:2rdm}
\end{figure}

In the main text, following \cite{Wietek2022}, we computed the two-body reduced density matrix by only considering the nearest-neighbor singlet pairing. For completeness, and to show that this is indeed the dominant type of pairing for our system, we compute the full two-body reduced density matrix (2RDM)
\begin{equation}
    \rho_2(\bm{r}_i\sigma_i,\bm{r}_j\sigma_j | \bm{r}_k\sigma_k, \bm{r}_l\sigma_l) = \langle \Delta^\dagger_{\bm{r}_i\sigma_i\bm{r}_j\sigma_j} \Delta_{\bm{r}_k\sigma_k, \bm{r}_l\sigma_l}\rangle,
\end{equation}
where 
\begin{equation}
    \Delta^\dagger_{\bm{r}_i\sigma_i\bm{r}_j\sigma_j} = c^\dagger_{\bm{r}_i\sigma_i}c^\dagger_{\bm{r}_j\sigma_j}.
\end{equation}
The spectrum $\varepsilon_n$ of $\rho_2$ is then defined as
\begin{gather}
    \rho_2(\bm{r}_i\sigma_i,\bm{r}_j\sigma_j | \bm{r}_k\sigma_k, \bm{r}_l\sigma_l) =\nonumber\\ \sum_n \varepsilon_n \chi_n^*(\bm{r}_i\sigma_i,\bm{r}_j\sigma_j)\chi_n(\bm{r}_k\sigma_k,\bm{r}_l\sigma_l)
\end{gather}
We compute this spectrum for a $16\times 6$ cylinder with $t'/t=0.2$, $\Phi=0$, and hole-doping $p=1/16$ (corresponding to $n_h=6$) in Fig.\ref{fig:2rdm}, removing as before elements where any two sites overlap, to exclude on-site density and spin correlations. 

We observe three dominant eigenvalues, indicating fragmented superconductivity with $n_h/2=3$ fragmented condensates. This is consistent with the picture obtained from the nearest-neighbour singlet assumption for the pairing matrix.

\bibliography{main}

\end{document}